\documentclass[12pt,preprint]{aastex}

\shorttitle{TESTING LAMBDA VIA SUPERNOVA COSMOGRAPHY}
\shortauthors{BOCHNER, PAPPAS, \& DONG}

\begin{document}

\title{Testing Lambda and the Limits of Cosmography\\ 
with the Union2.1 Supernova Compilation}

\author{Brett Bochner, Damon Pappas}
\affil{Department of Physics and Astronomy, Hofstra University, Hempstead, NY 11549, USA}
\email{brett\_bochner@alum.mit.edu, Brett.D.Bochner@hofstra.edu, Damon.A.Pappas@hofstra.edu}
\author{Menglu Dong}
\email{matilda.dong0101@gmail.com}

\begin{abstract}
We present a cosmographic study designed to test the simplest type 
of accelerating cosmology: a flat universe with matter and a 
cosmological constant ($\Lambda$). Hubble series expansions are fit 
to the SCP Union2.1 supernova data set to estimate the Hubble Constant 
($H_{0}$), the deceleration parameter ($q_{0}$), and the jerk parameter 
($j_{0}$). Flat $\Lambda$CDM models always require $j_{0} = 1$, 
providing a single-parameter test of the entire paradigm. 
Because of convergence issues for $z \gtrsim 1$, we focus on expansions 
using the newer redshift variable $y$; and to estimate the effects 
of ``model-building uncertainties'' -- the dependence of the output 
results upon the fitting function and parameters used -- we perform 
fits using five different distance indicator functions, and 
four different polynomial orders. We find that one cannot yet 
use the supernova data to reliably obtain more than four 
cosmological parameters; and that cosmographic estimates 
of $j_{0}$ remain dominated by model-building uncertainties, 
in conjunction with statistical and other error sources. 
While $j_{0} = 1$ remains consistent with Union2.1, the 
most restrictive bound that we can place is $j_{0} \sim [-7.6,8.5]$. 
To test the future prospects of cosmography with new standard candle 
data, ensembles of mock supernova data sets are created; and it is 
found that the best way to reduce model-building uncertainties on 
lower-order Hubble parameters (such as $\{ H_{0}, q_{0}, j_{0} \}$) 
is by limiting the redshift range of the data. Thus more and better 
$z \lesssim 1$ data, not higher-redshift data, is needed to 
sharpen cosmographic tests of flat $\Lambda$CDM.
\end{abstract}

\keywords{cosmological parameters --- cosmology: observations --- dark energy}
%
%\pacs{98.80.-k, 98.80.Es, 95.36.+x}

\section{\label{SecIntro}Introduction and Motivation}

The discovery of the acceleration of the universe 
\citep{RiessAccel98,PerlAccel99} solved certain major cosmological 
problems, such as the ``Age Crisis,'' while reconciling the low density 
of matter (as inferred from structure formation) with the fact 
of overall spatial flatness \citep{TurnerCosmoSense}. As a consequence, 
however, it created a new, fundamental problem: the entirely 
unknown nature of the force, effect, or substance responsible 
for this observed cosmic acceleration. 

Many different approaches have been taken toward explaining this 
(real or apparent) acceleration; and there is a voluminous literature on 
the different paradigms considered, which include modified gravity 
\citep[see][for a review]{CliftModGravReview}, 
inhomogeneity-perturbed observational effects 
\citep[e.g.,][]{KantowSwiss69,KantowSwiss98,CelerSchneid98,TomitaSNeVoid,
AlAmGronVoidAccel,ChungRomanLTB,GarfinkleLTBDE,BisNotLTBneglig,BolejkoSNeLTB}, 
and structure-formation-induced ``backreaction'' 
\citep[e.g.,][and references therein]{RefBackreactZeil,RefCausBackIJMPD2013}. 
In general, however, the standard approach usually involves the introduction 
of ``dark energy'' \citep{DEcoining}, a hypothesized substance which possesses 
negative pressure to drive the acceleration, yet which must remain largely 
unclumped in order to avoid a conflict with structure formation models. 
But whether or not a distinct dark energy substance turns out to be the 
correct culprit, determining the true nature of this mysterious effect 
represents an exceptionally difficult observational challenge. 

The question of whether or not dark energy exists in the form of a 
cosmological constant ($\Lambda$) is an extremely consequential one. The 
predominance of $\Lambda$ in the universe would lead to two well-known 
fine-tuning problems: one being the ``Cosmological Constant Problem'' 
\citep{KolbTurner}, relating to its magnitude being nonzero but far below 
the Planck (or any `natural') scale; and the other being the 
``Coincidence Problem'' \citep[e.g.,][]{ArkaniHamedCoinc}, the 
question of why $\Omega _{\Lambda} \sim O[1] \cdot \Omega _{M}$ in the 
current epoch (just in time to be seen by observers like ourselves), rather 
than being unmeasurably small or fatally large. Furthermore, a study of 
virialization with dark energy \citep{MaorLahavVirial} shows that $\Lambda$ is 
not even on the continuum of perfect fluids with general $w(z)$, but instead 
is a uniquely distinct entity. Thus the case for (or against) $\Lambda$ is a 
question with far-reaching implications.

To determine the physical nature of the dark energy -- i.e., its equation 
of state (EoS), $w(z) \equiv P / \rho$ -- one must place constraints up to 
(at least) the third-order term in the luminosity distance series expansions, 
since the first two terms tell us only about the present-day expansion rate 
(the Hubble Constant, $H_{0}$), and the effective amount of dark energy 
acting now ($q_{0} \sim w^{\mathrm{DE}}_{z = 0} \cdot \Omega _{\mathrm{DE}}$). 
Measuring its detailed, time-evolving behavior requires information 
beyond those first two terms, and so is very 
difficult to estimate. The issue of determining how many accurately estimated 
parameters can be obtained from cosmological data sets -- and how best to 
obtain them -- has been the subject of detailed analyses 
\citep[e.g.,][]{LinderHutLimParams,HutPCsystSNLS}, which generally highlight 
the difficulty of measuring $w(z)$. 

It is a common practice \citep[e.g.,][]{SuzRefUnion2pt1SNe} 
to combine a variety of different, complementary data sets -- such as 
compilations of Type Ia Supernovae (SNe), Cosmic Microwave Background (CMB) 
maps, standard ruler measurements from Baryon Acoustic Oscillations, 
and so on -- to obtain constraints on $w(z)$. But of all the different types 
of cosmological data sets, SN Ia standard candles 
are the only data which {\it directly} and continuously trace out the 
cosmic expansion as it evolved in the `recent' (moderate-redshift) 
universe, as the cosmic acceleration ``took over''; as a consequence, 
though beset by systematics \citep{HutPCsystSNLS} and large 
scatter, they are clearly the data most naturally suited for evaluating 
the cosmologically recent onset of acceleration, and for detailing its 
precise time evolution. Furthermore, in the 9-Year WMAP CMB data release, 
it was specifically the supernova data set which dramatically drove the 
combined analysis toward the $\Lambda$ condition of 
$w(z) \simeq w_{0} \simeq -1$ \citep[see Figure 10 of][]{WMAP9YrCosmoParams}; 
and from the Planck 2013 CMB results, it is similarly clear (from their 
comparisons of different SNe compilations) that the preferred 
parameter space for the dark energy EoS is particularly sensitive 
\citep[see Figure 36 of][]{Planck2013CosmoParams} to the choice of SNe data 
set used. It is therefore crucial that we properly interpret the `story' 
being told by the best available SNe data; and so, one goal of this paper 
is to illustrate what the most critically relevant data set -- SN Ia 
standard candles -- {\it can} and {\it cannot} yet prove 
about the cosmological constant. In particular, we focus here 
upon the comprehensive and homogeneously analyzed supernova data set 
known as the SCP Union2.1 SNe compilation \citep{SuzRefUnion2pt1SNe}. 

Now, if one's primary question is to determine, ``What kind of 
dark energy is accelerating the universe?,'' then the 
natural approach would be to evaluate time-varying EoS models 
for $w(z)$ by adopting some preferred parameterization -- such as 
``CPL'' \citep{ChevPolarCPL,LinderCPLexpFormalism}, 
$w(z) = w_{0} + w_{a} [z/(1+z)]$ --  
and fitting those parameters ($w_{0}, w_{a}$) to the SNe data. 
But though commonly employed, this method has certain disadvantages. 

First, the negative-pressure properties of dark energy (DE) imply that 
its density (and cosmological influence) were significantly lower at 
high redshift, making $w(z)$ an increasingly limp probe of the evolution 
of the acceleration as $z$ increases. Second, specifically testing the 
cosmological constant requires one to constrain the two-dimensional 
phase space of $(w_{0},w_{a})$, to simultaneously verify two conditions 
for $\Lambda$, $w_{0} = -1$ and $w_{a} = 0$; but given the limited 
information content of the data, a one-parameter test may be preferable, 
which can be done via the ``jerk parameter,'' $j_{0}$, related to the 
third derivative of the scale factor $a(t)$. Lastly, the most 
fundamental information being sought is not necessarily about what the 
dark energy (if it exists) is doing, but about what the {\it Universe} 
is doing. As noted in \citet[][]{RiessDiffParamsPriors}, the assumption 
of ``a simple dark-energy parameterization [like CPL] {\it is equivalent} 
[their emphasis] to a strong and unjustified prior on the nature of 
dark energy.'' Therefore, in the interest of placing all of the 
alternative paradigms (DE, backreaction, local voids, etc.) on an 
equal footing, rather than assuming {\it any} DE EoS function $w(z)$ 
at all, we instead use the kinematic approach known as ``cosmography'' 
\citep[e.g.,][]{CattVisserDfParamDfFits,CattVisserYvsZconverge,
CattVisserCosmographSNeFits}, to directly extract the Hubble series 
from the data without employing any prior assumptions about the physics 
underlying the cosmic expansion history. 

In the cosmographic method, one expands the luminosity distance (and/or 
related distance scale functions) in Taylor series using some choice of 
redshift parameter. The first three series terms introduce, respectfully: 
the Hubble Constant, $H_{0} \equiv \dot{a}_{0}/a_{0}$; the deceleration 
parameter, $q_{0} \equiv - (\ddot{a}_{0} / a_{0}) H_{0}^{-2}$ (which can be 
translated into the effective {\it amount} of ``dark energy,'' if desired); 
and the jerk parameter, $j_{0} \equiv (\dot{\ddot{a}}_{0} / a_{0}) H_{0}^{-3}$ 
(which can be translated into information about the {\it evolution} of the 
dark energy EoS -- e.g., $w_{a}$ -- if desired). A key advantage for cosmography 
in testing the hypothesis of a spatially flat universe containing only 
matter and $\Lambda$, is that such models must have $j(z) \equiv j_{0} = 1$ 
for all time \citep{CosmicJerkEqualsOne} -- regardless of the relative 
amounts of $\Omega _{\Lambda}$ and $\Omega _{M}$ -- as long as 
$\Omega _{\Lambda} + \Omega _{M} = 1$. Thus the entire hypothesis of 
flat $\Lambda$CDM can be falsified simply by searching for any deviation 
from $j_{0} = 1$. Thus $q_{0}$ (like $H_{0}$) becomes a ``nuisance parameter'' 
in this context, and the problem is indeed reduced to a search through 
a one-dimensional parameter space.

Cosmographic testing of $j_{0}$ via SNe data fitting has been done before, 
though with frustratingly discordant results. Several recent tests have used 
the SCP Union2 \citep{AmanRubinSCPunion2} supernova compilation (consisting 
of 557 SNe passing all data cuts, 23 fewer than Union2.1), in combination 
with different selections of auxiliary data sets being chosen for different 
analyses. For example, \citet{XuWangUnion2dAcosmograph} used Union2 SNe data 
(plus GRB and observational $H(z)$ data, etc.) to obtain a $\Lambda$-consistent 
(but fairly weak) constraint of approximately $j_{0} \simeq -5 \pm 7$. 
Similarly, \citet{CaiTuoj0Vals} obtained SNe-only estimates (using two 
different redshift expansions) of $j_{0} \simeq -1.83^{+5.85}_{-4.79}$ 
and $j_{0} \simeq -6.56^{+11.12}_{-21.40}$. 
\citet{VitagJerkSnapCrackleMG13} used different data set combinations 
(all in conjunction with Union2 SNe) for fitting luminosity distance curves, 
to obtain a variety of results from $j_{0} \simeq -7$ all the way to 
$j_{0} \simeq 5$, with the $\Lambda$-required result of $j_{0} \equiv 1$ 
being within the uncertainties for some of the fitted data combinations, 
but lying far outside the error bars for others. \citet{DemianskiHighZcosmog} 
found low $j_{0}$ values, including results more than 2$\sigma$ away from 
unity (though within 3$\sigma$), with variations of approximately 
$j_{0} \simeq 0.1 - 0.9$ for their different fits. This tendency of 
extreme variability of the $j_{0}$ results for different fitting functions, 
expansion variables, and/or data sets has typically been a hallmark of 
these and other cosmographic studies 
\citep[e.g.,][]{CapozzMCcosmograph,GruberUn2Cosmograph} 
done with this (and prior) SNe data compilations. 

The important point to take away from these discrepant results, is not 
that one must strive to find the ``right'' fitting function, variables, 
or data sets -- but that there {\it is no} absolutely ``right'' set of 
parameters. Instead, we must understand {\it why} different cosmographic 
fitting models give such differing results. The essential difficulty was 
explained and quantified (though with older SNe data) in a landmark study by 
\citet{CattVisserDfParamDfFits,CattVisserYvsZconverge,
CattVisserCosmographSNeFits}, in which they showed how the variability of 
parameter estimates is due to Hubble series truncation, because of the small 
number of terms that could be reliably estimated from the data: fewer fitting 
parameters leads to greater ``model-building uncertainty'' (referring to the 
variability of the output results, when different fitting models are used); but 
more fitting parameters (e.g., $H_{0}$, $q_{0}$, $j_{0}$, $s_{0}$, $c_{0}, \dots$) 
leads to greater statistical uncertainty in the best-fit values of each term 
(thus leading to unphysical parameter estimates, consistent statistically 
only because of their huge error bars). Our results in this research verify that 
intuition. Thus one cannot do better than to find the `sweet spot' which 
balances the model-building uncertainties on $H_{0}$, $q_{0}$, $j_{0}, \dots$ 
due to series truncation (by not using {\it too few} best-fit model parameters), 
versus limiting the statistical uncertainties on those parameters (by not using 
{\it too many} parameters), given the limited statistical power of the SNe data 
in hand. 

While such ``model-building'' uncertainties have been viewed somewhat mysteriously 
\citep[e.g.,][]{GruberUn2Cosmograph}, their origin is simply due to the limitation 
of having to fit a complicated (measured flux) curve with a Taylor series that has 
too few terms. For example, consider trying to fit a parabola of data 
(i.e., $y = a z + b z^{2}$) with a linear theory (i.e., $y = m z$). 
There is no uniquely correct way of doing this; and by variously emphasizing 
either the low-$z$ or high-$z$ data (e.g., via multiplication by various powers 
of $(1+z)$), one will get different `best-fit' slopes $m$. To fix this problem, 
there are two straightforward ways of getting a uniquely determined answer: 
either add more polynomial terms to the Taylor expansion fit (i.e., 
$y = m_{1} z + m_{2} z^{2} + \dots$), or cut off the data to leave only 
a small range in $z$, so that the `curve' of the parabola does not manifest itself.  
Unfortunately, as our results prove, though increasing the number of best-fit 
Taylor series coefficients from 3 (the minimum needed for estimating $j_{0}$) 
to 6 does indeed reduce the model-building uncertainties to very small levels, 
the act of including more fitting terms $m_{i}$ also does lead to a dramatic 
increase in the statistical uncertainties on {\it all} of the best-fit 
coefficients, including the lower-order ones actually used to calculate 
$H_{0}$, $q_{0}$, and $j_{0}$. Similarly, cutting off the SNe data above some 
high-$z$ threshold {\it also} increases the statistical uncertainties, 
despite successfully reducing the model-building uncertainties. Thus all that 
one can strive for, is to find a balance between the statistical versus the 
model-building uncertainties, in order to determine what is 
``the best that one can do'' in a cosmographic analysis of a given data set.

This problem of parameter fitting indeterminacy is {\it not} unique to 
cosmographic methods, but also occurs with fits using dark energy EoS 
functions with a very limited number of best-fit coefficients -- 
as demonstrated, for example, in \citet{RiessDiffParamsPriors}, and also by 
the ``Mirage of $w = -1$'' problem \citep{LinderWneg1Mirage} for fits with 
$w(z)$ models limited to a single, constant value (i.e., $w(z) \equiv w_{0}$). 
No type of theoretical model is immune to model-building uncertainties, 
if the number of adjustable parameters used in the fitting process is 
insufficient to model the curve followed by the data.

The core of the aforementioned study 
\citep{CattVisserDfParamDfFits,CattVisserYvsZconverge,CattVisserCosmographSNeFits} 
involved a comprehensive collection of fits using five different 
distance indicator functions: the traditional luminosity 
distance $d_\mathrm{L} (z)$, along with four other functions related 
to $d_\mathrm{L}$ by different powers of ($1+z$). And considering 
series convergence issues for $z \gtrsim 1$, they also fit the data 
to series expansions in ``$y$-redshift'' variable, $y \equiv z/(1+z)$. 
Our work here builds upon and expands their work, utilizing newer SNe data 
and higher-order polynomial fitting functions. To that end, we have 
conducted a systematically designed study of the SCP Union2.1 data set 
(a compilation of 580 SNe Ia passing all data cuts), with a full set 
of best-fit functions that includes: (i) five different 
distance indicator functions; (ii) expansions done using both $z$-redshift 
and $y$-redshift; and, (iii) Hubble series fits using several different 
polynomial orders (with functions possessing $3$, $4$, $5$, and $6$ 
fitted parameters, respectively). A fully comprehensive discussion of all of 
these simulations and their implications is available in a separate preprint 
\citep{BochnerPDLamDev}, henceforth BPDv2; here we provide an overview of 
our methodology, and a discussion of the key results.

From our overall set of $5 \times 2 \times 4 = 40$ fits to the Union2.1 
compilation, it is clear that the value of the jerk parameter $j_{0}$ cannot 
yet be narrowed down with great precision, even using Union2.1 as a virtually 
continuous tracer over most of the acceleration epoch. A useful quote of 
our constraints (see Section~\ref{SecCosmogResults}) results in a range 
for the jerk parameter of $j_{0} \sim [-7.6,8.5]$ -- which, though being 
completely consistent with a cosmological constant ($j_{0} \equiv 1$), 
still remains far from providing a stringent test of $\Lambda$ versus 
the more dynamical forms of dark energy. Finally, two classes of mock data 
sets are constructed (involving 200 randomized simulations for each case), 
which are evaluated to determine how the current constraints may be improved 
with future supernova data.

\section{\label{SecCosmogMethods}Cosmographic Methodology}

Due to the expansion of the universe, the definition of distance 
in cosmology (between an early emitter and a later observer) 
is fundamentally ambiguous. For example, the ``luminosity distance'' 
$d_\mathrm{L}(z)$ is typically defined from the relationship between 
the known luminosity $L$ for a standard candle (e.g., a SN Ia), 
and its measured flux $F$, via: $F = L / (4 \pi d_\mathrm{L}^2)$. 
In FLRW models, this definition sets 
$d_\mathrm{L} \equiv a_{0} r (1 + z)$, where $a_{0}$ is the 
present-day scale factor and $r \equiv r(z)$ is the comoving 
coordinate distance to the emitting object. Alternatively, the 
``angular diameter distance'' $d_\mathrm{A}(z)$ is typically defined 
from the relationship between the measured angular size $\theta$ 
of a ``standard ruler,'' and its known physical diameter $D$, as: 
$\theta = D / d_\mathrm{A}$. This definition sets 
$d_\mathrm{A} \equiv a_{0} r / (1 + z)$, and hence, 
$d_\mathrm{A} = d_\mathrm{L} (1+z)^{-2}$. There is, however, no end 
to the different types of distance indicator functions which one may introduce; 
for example \citep[see][]{CattVisserDfParamDfFits}, one could also consider 
the ``photon flux distance,'' $d_\mathrm{F} = d_\mathrm{L} (1+z)^{-1/2}$; 
the ``photon count distance,'' $d_\mathrm{P} = d_\mathrm{L} (1+z)^{-1}$; 
the ``deceleration distance,'' $d_\mathrm{Q} = d_\mathrm{L} (1+z)^{-3/2}$; 
and/or any other distance indicator function related by various powers 
or functions of $(1+z)$. 

Despite their evocative names (and idiosyncratic origins), all of these 
distance scale functions are equally good expressions to use for fitting to 
any type of cosmological data set. Other than traditional usage, there is 
no particular reason why one must (or should) fit flux data to the 
``luminosity distance'' function, $d_\mathrm{L} \propto F^{-1/2}$, as opposed 
to any other function of flux $F$ (and redshift $z$). The key realization by 
\citet{CattVisserDfParamDfFits,CattVisserYvsZconverge,CattVisserCosmographSNeFits}, 
was that the use of a {\it different} distance function to fit the {\it same} 
data ends up yielding radically different results. In particular, the best-fit 
value of a particular cosmological parameter will be changed, from one 
distance scale function to the next, in an evenly spaced manner (which they 
explained theoretically as an effect of series truncation, resulting from 
the unavoidably finite number of best-fit series coefficients). This same 
behavior is also seen in our results here, and so the error budget to be 
minimized {\it must} include these model-building uncertainties, in tandem 
with the statistical uncertainties on the estimated cosmological parameters. 

Complete expressions for each Hubble series for 
$\{d_\mathrm{L}(z),d_\mathrm{F}(z),d_\mathrm{P}(z),
d_\mathrm{Q}(z),d_\mathrm{A}(z)\}$, and for 
$\{d_\mathrm{L}(y), \dots , d_\mathrm{A}(y)\}$ -- up to three polynomial 
terms, the minimum number required to estimate $j_{0}$ -- have been given 
in \citet[][and their many related references]{CattVisserDfParamDfFits}. 
Furthermore, series expressions are also are given for these functions in 
the alternative form, $\ln \{ [ d_\mathrm{L}(z) ] / z \}$, etc., which 
conveniently removes a varying but cosmologically irrelevant term from the 
plotted and fitted functions. Due to this and other advantages, we have 
performed fits (BPDv2) using their ten different series expansions for: 
$\ln \{ [ d_\mathrm{L}(z) ] / z \}, \dots , \ln \{ [ d_\mathrm{A}(z) ] / z \}$, 
and: 
$\ln \{ [ d_\mathrm{L}(y) ] / y \}, \dots , \ln \{ [ d_\mathrm{A}(y) ] / y \}$. 
Due to the unavoidable series convergence concerns for $z$-redshift, however, 
in this paper we focus exclusively on the fits performed using 
the $y$-redshift variable.

As one example case to describe explicitly, consider the expansion: 
\begin{eqnarray}
\ln \{ [ d_\mathrm{P}(y) ] / y \} & = & 
\ln (\frac{c}{H_{0}}) 
+ \frac{1}{2} (1 - q_{0}) y 
\nonumber 
\\ 
& + & \frac{1}{24} (5 - 2 q_{0} + 9 q_{0} ^{2} - 4 j_{0}) y^{2} 
+ O[y^{3}] , 
\label{dPdyExpansion}
\end{eqnarray}
where we have already assumed flatness everywhere (i.e., setting terms 
like $(j_{0} + \Omega _{0})$ equal to $(j_{0} + 1)$ in all expressions). 

The Union2.1 compilation data is described in \citet{SuzRefUnion2pt1SNe}, 
and the data itself is publicly available 
from the Supernova Cosmology Project website, 
{\it http://supernova.lbl.gov/Union}. Their ``Magnitude vs.\ Redshift 
Table'' lists $(z,\mu,\sigma _{\mu})$ for each SN Ia, where $\mu$ 
represents the ``distance modulus'' data, and the $\sigma _{\mu}$ are 
their (uncorrelated) statistical errors. The $z$-redshift values can be 
converted into $y$-redshift as discussed above, and each of the 
$(\mu, \sigma _{\mu})$ values can be trivially converted into data 
and uncertainty values in any form required -- e.g., 
$\ln ( d_\mathrm{P} / y )$ -- for comparison to expansions such as 
Equation~\ref{dPdyExpansion}. 

Now consider a fitting polynomial of the form: 
\begin{equation}
F(y) \equiv p_{0} + p_{1} y + p_{2} y^{2} + \cdots + p_{n} y^{n} ~. 
\label{yPolyExpansion}
\end{equation}
This function, going up to order $y^{n}$, will possess $N \equiv (n+1)$ 
optimizable parameters. We need to use {\it at least} $N = 3$ 
polynomial terms in order to estimate $j_{0}$, though in principle one can 
include any number of additional high-order series terms. Though the 
formulas for $\{ H_{0},q_{0},j_{0} \}$ -- derived from the first three terms 
of expressions like Equation~\ref{dPdyExpansion} -- do not change 
based on $N$, we will see that their best-fit values do in fact depend 
significantly upon $N$, due to changing statistical and model-building 
uncertainties. Our modeling includes the results from fits using 
$N = \{ 3, 4, 5, 6 \}$ (for each redshift variable and distance scale function), 
evaluating each fit in light of the $F$-test of additional terms 
\citep{CattVisserDfParamDfFits,CattVisserYvsZconverge,CattVisserCosmographSNeFits}. 

When fit to a SNe data set with a total of $N_{\textrm{SN}}$ supernovae 
(where $N_{\textrm{SN}} = 580$ for Union2.1), a polynomial $F(y)$ 
with $N$ terms will result in a best-fit with 
$(N_{\textrm{SN}} - N)$ degrees of freedom. We then compute: 
\begin{equation}
\chi ^{2} (p_{0},p_{1}, \dots, p_{n}) 
= \sum ^{580}_{i} \frac{[d_{i} - F(y_{i})]^{2}}{\sigma _{i} ^{2}} ~, 
\label{LstSqrsEqn}
\end{equation}
where $(d _{i},\sigma _{i})$ are the data point and statistical uncertainty 
for the $i^\mathrm{th}$ supernova. This $\chi ^{2}$ is then minimized 
in order to determine the best-fit values and sigmas of 
$(p_{0},p_{1}, \dots, p_{n}) \equiv \vec{p}$, for this form of fitting function. 

The optimization is done using standard statistical techniques. The elements 
of the coefficient matrix $\textbf{\textit{A}} \equiv \{ a_{jk} \}$ 
(and an auxiliary vector $\vec{b} \equiv \{ b_{k} \} $) 
are calculated via: 
\begin{equation}
a_{jk} = \sum ^{580}_{i} \frac{(x_{i})^{j} (x_{i})^{k}}{\sigma _{i} ^{2}} ~,~ 
b_{k} = \sum ^{580}_{i} \frac{(d_{i}) (x_{i})^{k}}{\sigma _{i} ^{2}} ~, 
\label{CoeffMatCalc}
\end{equation}
where $x_{i}$ is the $i^\mathrm{th}$ redshift value 
(either $z_{i}$ or $y_{i}$), and $(j,k)$ run through $\{ 0,n \}$. 
Then with the parameter error covariance matrix 
$\textbf{\textit{Z}} \equiv \{ z_{jk} \} = \textbf{\textit{A}}^{-1}$, 
we get the optimized parameters as $\vec{p} = \textbf{\textit{Z}} \vec{b}$, 
with their sigmas obtained from the diagonal elements of $\textbf{\textit{Z}}$ 
(i.e., $\sigma _{p_{k}} = \sqrt{z_{kk}}$).

Once these best-fit polynomial coefficients have been determined, 
we can invert the appropriate Hubble series to obtain the optimized 
cosmological parameters corresponding to them. In cosmographic expansions 
like Equation~\ref{dPdyExpansion} for the various distance scale functions, 
each new series term introduces one new cosmological parameter; and our 
results show that the uncertainties typically get significantly larger 
for each higher-order coefficient 
(i.e., $\sigma _{p_{0}} \ll \sigma _{p_{1}} \ll \sigma _{p_{2}} 
\ll \sigma _{p_{3}} \dots $, is usually true). It is therefore 
a good procedure to obtain $H_{0}$ (and $\sigma _{H_{0}}$) solely from $p_{0}$ 
(in fact, the Hubble constant {\it only} appears in $p_{0}$), and to get 
$q_{0}$ solely from $p_{1}$; we also get $j_{0}$ solely from $p_{2}$ 
(regardless of the total number of expansion terms, $N \ge 3$), though with 
some dependence on $q_{0}$ (and thus $p_{1}$). 

For the example discussed here ($\ln \{ [ d_\mathrm{P}(y) ] / y \}$), 
we compare Equations~\ref{dPdyExpansion} and \ref{yPolyExpansion} to obtain: 
\begin{equation}
H_{0} = c e^{- p_{0}} ~, 
\label{CosmoParamH0Eqn}
\end{equation}
\begin{equation}
q_{0} = 1 - 2 p_{1} ~, 
\label{CosmoParamq0Eqn}
\end{equation}
\begin{equation}
j_{0} = -6 p_{2} + \frac{1}{4} (5 - 2 q_{0} + 9 q_{0}^2) ~.
\label{CosmoParamj0Eqn}
\end{equation}
Additionally, from the Friedmann Robertson-Walker acceleration equation, 
plus the definition of $q_{0}$, one obtains $q_{0} = (1 + 3 w_{0})/2$, 
which can be inverted to give: 
\begin{equation}
w_{0} = \frac{1}{3} (2 q_{0} - 1) ~. 
\label{w0fromq0eqn}
\end{equation}
Note, though, that this EoS, $w_{0} \equiv w^{\mathrm{Obs}}_{0}$, 
does {\it not} represent the equation of state of any ``dark energy'' 
component by itself, but rather represents 
the effective {\it total} EoS of all of the cosmic contents averaged 
together, as inferred from observations of the overall evolution. 

If one wishes, it is not difficult to relate these directly observed 
(i.e., cosmographic) parameters, 
$(w^{\mathrm{Obs}}_{0},j^{\mathrm{Obs}}_{0})$, to the EoS 
parameters of an underlying dark energy model with a dynamical equation 
of state $w^{\mathrm{DE}}(z)$, given some particular parameterization. 
Specifically, using the popular CPL parameterization, 
$w^{\mathrm{DE}}(z) 
= w^{\mathrm{DE}}_{0} + w^{\mathrm{DE}}_{\mathrm{a}} [z/(1+z)] 
= w^{\mathrm{DE}}_{0} + w^{\mathrm{DE}}_{\mathrm{a}} y 
\equiv w^{\mathrm{DE}}(y)$, 
along with the assumption of spatial flatness, 
one obtains \citep{BochnerAccelPaperI}: 
\begin{equation} 
w^\mathrm{Obs}_{0} = \Omega _{\mathrm{DE}} ~ w^{\mathrm{DE}}_{0} ~ , 
\label{Eqnw0FnOfw0wa}
\end{equation}
which makes obvious sense; and:
\begin{equation} 
j^\mathrm{Obs}_{0} = \{ 1 + [\frac{9}{2} 
\Omega _{\mathrm{DE}} w^{\mathrm{DE}}_{0} (1 + w^{\mathrm{DE}}_{0})] + 
[\frac{3}{2} \Omega _{\mathrm{DE}} w^{\mathrm{DE}}_{\mathrm{a}}] \} 
 ~ , 
\label{Eqnj0FnOfw0wa}
\end{equation} 
which gives $j^\mathrm{Obs}_{0} = 1$ for the 
cosmological constant case (i.e., $w^{\mathrm{DE}}_{0} = -1 , 
w^{\mathrm{DE}}_{\mathrm{a}} = 0$), for {\it any} value 
of $\Omega _{\mathrm{DE}}$, just as required. But while these formulas 
allow one to convert our cosmographic parameter results into 
dark energy EoS parameters, we note again that since the two cosmographic 
parameters, $(w^\mathrm{Obs}_{0},j^\mathrm{Obs}_{0})$, are divided up 
(due to observational degeneracies) into {\it three} DE parameters, 
$(\Omega _{\mathrm{DE}},w^{\mathrm{DE}}_{0},w^{\mathrm{DE}}_{\mathrm{a}})$, 
one requires some auxiliary condition to be imposed among them in order to 
convert cosmographic constraints into dark energy constraints. 
We have therefore placed the translation of our Union2.1 cosmographic 
results into CPL parameter constraints in an appendix of BPDv2.

The parameter uncertainty values, 
$\{ \sigma _{H_{0}},\sigma _{q_{0}},\sigma _{w_{0}} \}$, can be 
computed (using elementary error propagation) from, respectively, 
Equations~\ref{CosmoParamH0Eqn},\ref{CosmoParamq0Eqn}, 
and \ref{w0fromq0eqn} (and from similar equations for each of the 
various distance scale expansions). Since $H_{0}$ is drawn solely 
from $p_{0}$, and $\{ q_{0},w_{0} \}$ are drawn solely from $p_{1}$, 
there are no issues of parameter covariance for them once the best-fit 
values of $\{ p_{i} \}$ have been determined. But since $j_{0}$ depends 
upon both $p_{1}$ (via $q_{0}$) and $p_{2}$, the relevant terms 
in the parameter covariance matrix must be considered. 
For example (for $\ln \{ [ d_\mathrm{P}(y) ] / y \}$), by re-writing 
Equation~\ref{CosmoParamj0Eqn} as $j_{0} \equiv F(p_{1},p_{2})$, 
we obtain:
\begin{equation} 
\sigma _{j_{0}} = \{ 
[ (\frac{\partial F}{\partial p_{1}}) ^2 \cdot z_{11} ] 
+ [ (\frac{\partial F}{\partial p_{2}}) ^2 \cdot z_{22} ] 
+ [ 2 (\frac{\partial F}{\partial p_{1}}) 
(\frac{\partial F} {\partial p_{2}}) \cdot z_{12} ] \} ^{1/2}
 ~ , 
\label{EqnSigj0withCov}
\end{equation} 
with a similar $\sigma _{j_{0}}$ calculation being done for 
each different fitting function used. 

Lastly, we note the obvious fact that the parameter estimation strategy 
outlined here is a considerably simplified version of the fitting process 
often used \citep[e.g.,][]{SalzanoRefRecFits} for obtaining 
state-of-the-art cosmological parameter constraints from SNe Ia. 
This is done deliberately, since more important (in this paper) 
than providing the most comprehensive possible constraints, is our goal 
of exploring the intricacies of the use of cosmography in standard candle 
data fitting, demonstrating its advantages and limitations independently 
of any particular survey or its systematics.

\section{\label{SecCosmogResults}Cosmological Parameter Ranges 
Resulting from the Cosmographic Fits}

\begin{figure}
\includegraphics[width=\linewidth]{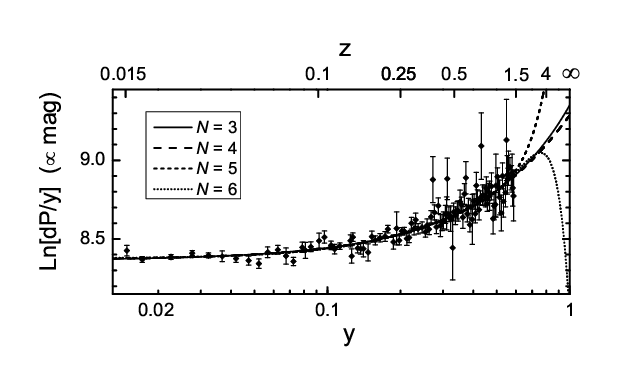}
\caption{SN Ia data from the SCP Union2.1 compilation, 
binned and averaged (with $\Delta y = 0.005$). The data is presented 
in the form $\ln ( d_\mathrm{P} / y )$ and fit with four different 
polynomials $F(y)$ (as per Equation~\ref{yPolyExpansion}), 
with $N = \{ 3, 4, 5, 6 \}$ optimized polynomial coefficients, 
respectively. At the high-redshift end, the $N = 5$ case runs away 
to larger (finite) values, while $N = 6$ turns and drops downward; 
but the $N = 3$ and $N = 4$ fits change very slowly and 
remain close together as $y \rightarrow 1$, $z \rightarrow \infty$.
\label{Fig1BPDrev2FitLNdPdivY}}
\end{figure}

First, examining the behavior of the cosmographic fits themselves,
consider the example given in Figure~\ref{Fig1BPDrev2FitLNdPdivY}, 
showing the SNe data in terms of photon count distance, 
$\ln ( d_\mathrm{P} / y )$, plotted against its four relevant 
$y$-redshift polynomial fits. The residuals (not shown) for these 
different fits are very similar to one another -- all fitting polynomials 
are very tightly constrained by the data below $z \lesssim 1.4$ -- 
and despite a handful of individual SNe outliers (each possessing 
large error bars), no obvious systematic trends are apparent.

From Figure~\ref{Fig1BPDrev2FitLNdPdivY}, it is clear that while 
all four fitting functions are virtually identical over the 
redshift range for which SNe data exists, the two higher-order 
polynomials ($N = 5$ and $N = 6$) become very poorly constrained 
as soon as the data runs out, for $y \gtrsim 0.59$ ($z \gtrsim 1.4$). 
This behavior foreshadows the results to be given below,  
that if one goes beyond the $N = 4$ case, the best-fit 
cosmological parameter values start changing drastically. 
Conversely, since higher-order polynomial terms are needed in 
order to retain accuracy of the Taylor series far from $y = 0$, 
the presence of higher-redshift data would actually {\it require} 
more series terms $N$ to be used in the cosmographic fitting process 
for the results to be reliable. 

The output cosmological parameters derived 
(via Equations like \ref{CosmoParamH0Eqn}-\ref{w0fromq0eqn}) 
from our cosmographic polynomial fits to the 
Union2.1 data set are listed in Table~\ref{TAByRedshiftCosParams} 
for the $y$-redshift expansions. These parameter estimation results 
are very much in line with our expectations from the prior studies 
by \citet{CattVisserDfParamDfFits,CattVisserYvsZconverge,
CattVisserCosmographSNeFits}. First, there is indeed a jump in each 
parameter value when going from one distance indicator function 
to the next, and the jumps are quite regularly spaced (within 
a given polynomial order $n$), just as expected. Furthermore, 
the variation in the results for different distance function fits 
is particularly large when fewer polynomial terms are used 
(i.e., smaller $N \equiv n + 1$), making the effects of 
model-building error due to series truncation very obvious.

\begin{table}
\begin{center}
\caption{\label{TAByRedshiftCosParams}
Cosmographic Parameters from Hubble Series Expansions in 
$y$-Redshift, fitted to the SCP Union2.1 Type Ia supernova 
data set ($N_{\mathrm{SNe}} = 580$).}
{
\begin{tabular}{cccccc}
Fitting Function & Fit Terms $N$ & $H_{0}$ & $q_{0}$ & $w_{0}$ & $j_{0}$ \\
\tableline
$\ln \{ [ d_\mathrm{L}(y) ] / y \}$ & 
     & 69.722 & -0.381 & -0.587 & -2.18 ($\pm 1.28$) \\
$\ln \{ [ d_\mathrm{F}(y) ] / y \}$ & 
     & 69.879 & -0.492 & -0.661 & -0.50 ($\pm 1.35$) \\
$\ln \{ [ d_\mathrm{P}(y) ] / y \}$ & 3 
     & 70.036 & -0.603 & -0.735 & 1.24 ($\pm 1.43$) \\
$\ln \{ [ d_\mathrm{Q}(y) ] / y \}$ & 
     & 70.194 & -0.714 & -0.809 & 3.04 ($\pm 1.50$) \\
$\ln \{ [ d_\mathrm{A}(y) ] / y \}$ & 
     & 70.352 & -0.825 & -0.883 & 4.89 ($\pm 1.58$) \\
 &      & ($\pm 0.509-0.514$) & ($\pm 0.154$) & ($\pm 0.103$) & ($\dots$) \\
\tableline
$\ln \{ [ d_\mathrm{L}(y) ] / y \}$ & 
     & 70.037 & -0.623 & -0.749 & 2.17 ($\pm 6.29$) \\
$\ln \{ [ d_\mathrm{F}(y) ] / y \}$ & 
     & 70.003 & -0.587 & -0.725 & 1.22 ($\pm 6.23$) \\
$\ln \{ [ d_\mathrm{P}(y) ] / y \}$ & 4 
     & 69.968 & -0.551 & -0.700 & 0.28 ($\pm 6.17$) \\
$\ln \{ [ d_\mathrm{Q}(y) ] / y \}$ & 
     & 69.933 & -0.514 & -0.676 & -0.65 ($\pm 6.11$) \\
$\ln \{ [ d_\mathrm{A}(y) ] / y \}$ & 
     & 69.899 & -0.478 & -0.652 & -1.58 ($\pm 6.05$) \\
 &      & ($\pm 0.668-0.669$) & ($\pm 0.365$) & ($\pm 0.243$) & ($\dots$) \\
\tableline
$\ln \{ [ d_\mathrm{L}(y) ] / y \}$ & 
     & 69.374 & 0.115 & -0.257 & -17.89 ($\pm 18.16$) \\
$\ln \{ [ d_\mathrm{F}(y) ] / y \}$ & 
     & 69.383 & 0.104 & -0.264 & -17.50 ($\pm 18.20$)  \\
$\ln \{ [ d_\mathrm{P}(y) ] / y \}$ & 5 
     & 69.391 & 0.092 & -0.272 & -17.11 ($\pm 18.24$)  \\
$\ln \{ [ d_\mathrm{Q}(y) ] / y \}$ & 
     & 69.399 & 0.081 & -0.279 & -16.71 ($\pm 18.28$)  \\
$\ln \{ [ d_\mathrm{A}(y) ] / y \}$ & 
     & 69.408 & 0.069 & -0.287 & -16.32 ($\pm 18.32$)  \\
 &      & ($\pm 0.901-0.902$) & ($\pm 0.775$) & ($\pm 0.516$) & ($\dots$) \\
\tableline
$\ln \{ [ d_\mathrm{L}(y) ] / y \}$ & 
     & 69.088 & 0.523 & 0.015 & -32.64 ($\pm 47.18$)  \\
$\ln \{ [ d_\mathrm{F}(y) ] / y \}$ & 
     & 69.086 & 0.526 & 0.018 & -32.80 ($\pm 47.16$)  \\
$\ln \{ [ d_\mathrm{P}(y) ] / y \}$ & 6 
     & 69.084 & 0.530 & 0.020 & -32.95 ($\pm 47.14$)  \\
$\ln \{ [ d_\mathrm{Q}(y) ] / y \}$ & 
     & 69.082 & 0.533 & 0.022 & -33.11 ($\pm 47.12$)  \\
$\ln \{ [ d_\mathrm{A}(y) ] / y \}$ & 
     & 69.080 & 0.537 & 0.024 & -33.26 ($\pm 47.09$)  \\
 &      & ($\pm 1.253-1.254$) & ($\pm 1.471$) & ($\pm 0.981$) & ($\dots$) \\
\tableline
\end{tabular}
}
\end{center}
\end{table}

For convenience in this discussion, we define the following 
terminology: let ``$C^{N}_\mathrm{X} (R)$'' represent the best-fit 
value of cosmological parameter ``$C_{0}$'' 
($C \in \{ H, q, w, j \}$), estimated with distance function 
$d_\mathrm{``X''}$ 
($\mathrm{X} \in \{\mathrm{L}, \mathrm{F}, \mathrm{P}, \mathrm{Q}, 
\mathrm{A} \}$), when expanded in redshift variable ``$R$'' 
($R \in \{z, y \}$), using $N$ terms in the polynomial expansion 
($N \in \{ 3,4,5,6 \}$). So for example, 
$H^{\mathrm{3}}_\mathrm{L} (y) = 69.722$ is the best-fit value of 
$H_{0}$ from the $N = 3$ SNe data fit done with 
$\ln \{ [ d_\mathrm{L}(y) ] / y \}$. Relatedly, let 
``$\Delta _\mathrm{LA} C ^{N} (R)$'' represent the 
{\it absolute net change} in a cosmological parameter when going 
from the distance function at one ``extreme'' (given our admittedly 
arbitrary choice to use 5 distance indicator functions here), 
to the function at the other extreme: for example, 
$\Delta _\mathrm{LA} H ^{3} (y) = 
\vert H^{3}_\mathrm{L} (y) - H^{3}_\mathrm{A} (y) \vert 
= \vert 69.722 - 70.352 \vert = 0.63$. 

From the numbers in Table~\ref{TAByRedshiftCosParams}, 
we see that the variation of the output parameters is crucially dependent 
upon the number of terms used in the fit. For example, 
$\Delta _\mathrm{LA} q ^{3} (y) = 0.444$, rendering the $N = 3$ case 
almost useless for getting precision measurements of the strength of 
the acceleration (and thus of the DE EoS, $w_{0}$); while the $N = 6$ 
case produces the dramatically more stable result of 
$\Delta _\mathrm{LA} q ^{6} (z) = 0.014$. And considering our interest 
in testing $\Lambda$ via precision measurements of $j_{0}$, 
we note that $\Delta _\mathrm{LA} j ^{6} (y) = 0.62$ (a reasonably 
small variation under the circumstances), but the $N = 3$ case yields 
$\Delta _\mathrm{LA} j ^{3} (y) = 7.07$, which is not nearly 
small enough for precision tests of $\Lambda$.

We are therefore strongly motivated to go beyond the $N = 3$ case, 
using as many optimizable parameters as possible. But how many terms 
can be usefully added? Despite the much smaller model-building uncertainty 
variations for the $N = 6$ case -- $\Delta _\mathrm{LA} w ^{6} (y) = 0.009$, 
as well as $\Delta _\mathrm{LA} j ^{6} (y) = 0.62$ -- the actual 
best-fit cosmological parameter values from the $N > 4$ cases 
make little physical sense. First, the values of $w^{5}_\mathrm{X} (y)$ 
and $w^{6}_\mathrm{X} (y)$ are all significantly greater than $(-1/3)$ -- 
i.e., $\{ q^{5}_\mathrm{X} (y), q^{6}_\mathrm{X} (y) \} > 0$ -- implying 
that the universe is {\it not even accelerating} according to the 
best-fit cosmic EoS from those fits. Furthermore, 
these cases also produce implausible best-fit estimates of the 
jerk parameter, with $j^{5}_\mathrm{X} (y) \sim -17$ to $-18$, 
and the $N = 6$ case going as far away from $\Lambda$ as 
$j^{6}_\mathrm{X} (y) \sim -33$. 
Recalling Equation~\ref{Eqnj0FnOfw0wa}, and assuming 
$\Omega _{\mathrm{DE}} \sim 0.7$, a jerk parameter as negative 
as just $j_{0} \lesssim -4$ leads already to a DE evolution 
as strong as $w^{\mathrm{DE}}_{\mathrm{a}} \lesssim -4.8$ 
(for $w^{\mathrm{DE}}_{0} \sim -1$). Compare that result 
to those from Table 7 of \citet{SuzRefUnion2pt1SNe}, which -- 
over all of their cases, and including both statistical 
{\it and} systematic error ranges -- gives a total estimated 
range of $w^{\mathrm{DE}}_{\mathrm{a}} \sim [-4.4,1.3]$. 
The best-fit parameters from our large-$N$ cosmographic fits are 
clearly very different from those quoted in such dynamical DE analyses. 

It is important to note, however, that our large-$N$ cosmographic 
results here are {\it not} statistically inconsistent with those cited 
values. The reason for this, is that increasing the number of 
optimizable polynomial terms in the model fits, without any 
additional information content in the data to constrain them, 
results in greatly increased statistical uncertainties for each of 
the parameters. Our apparently unusual cosmological estimates are 
therefore due to those unavoidably large parameter uncertainties. 
In the fits just discussed, {\it all} of the  
$w^{5}_\mathrm{X} (y)$ and $w^{6}_\mathrm{X} (y)$ values are 
actually within 1$\sigma$ of $w^\mathrm{Obs}_{0} = -0.7$, 
considering their large error bars; and similarly, for 
$j^{5}_\mathrm{X} (y)$ and $j^{6}_\mathrm{X} (y)$, 
8 out of the 10 cases are within 1$\sigma$ of the $j_{0} = 1$ 
required by flat $\Lambda$CDM, with the other 2 cases not being 
far away. Thus these $N = \{ 5, 6 \}$ fits, 
while being statistically consistent with previous results, 
and demonstrably model-independent -- i.e., effectively immune 
to model-building uncertainties (especially for $N = 6$) -- are 
limited to providing ``reliable'' but unimpressively broad 
cosmological constraints.  

This cosmographic study therefore reveals the dilemma when 
using such data to attempt to estimate the jerk parameter 
to within $\Delta  j_{0} \sim 1$ or so. If one sticks to $N = 3$, 
then one may quote (for example) 
$j_{0} = j^{3}_\mathrm{L} (y) = -0.95 \pm 0.14$ from the 
$\ln \{ [ d_\mathrm{L}(z) ] / z \}$ fit, which seems like a 
fairly tight constraint that strongly excludes $\Lambda$CDM. 
But this result is completely spurious, since if one instead quoted 
the fit to $\ln \{ [ d_\mathrm{A}(z) ] / z \}$, then the result would 
have the completely different value, 
$j_{0} = j^{3}_\mathrm{A} (y) = 2.38 \pm 0.24$. The mutual 
inconsistency of these numbers is due to the great sensitivity of the 
results to the fitting model chosen, when only three parameters 
are available for optimization. Thus model-building uncertainty 
completely trumps the statistical uncertainties quoted in such fits, 
and leads to a false confidence in the results for $j_{0}$. 
Going to $N = 6$, on the other hand -- which ensures the 
fitting-model-independence of the results (i.e., 
$\Delta _\mathrm{LA} j ^{6} (y) < 1$) --  increases the statistical 
uncertainties on $j_{0}$ so much ($\sigma _{j_{0}} \sim 47$), 
that the constraints on flat $\Lambda$CDM from the cosmographic fit 
are very robust but extremely weak. 

Note that this dilemma of having to balance precision versus accuracy 
is not due to the method of cosmography, per se -- though cosmographic 
fitting does make the problem extremely transparent -- but instead comes 
from the inherently limited information contained within the data set itself. 
Though one cannot draw an exact parallel between cosmography and the 
very different dynamical DE fits that start by assuming some EoS 
function $w(z)$, it is interesting to make some analogies. Our 
case with $N = 3$ optimizable parameters, $\{ H_{0},w_{0},j_{0} \}$ 
derived from $\{ p_{0},p_{1},p_{2} \}$, is arguably analogous to 
constant-$w$ models, with the three optimizable parameters: $H_{0}$, 
$\Omega _{\mathrm{DE}}$, and $w \equiv w^{\mathrm{DE}}_{0}$. Given the 
unacceptable sensitivity that we have found here for the $N = 3$ case 
to the particular fitting model chosen, in conjunction with the well-known 
``Mirage of $w = -1$'' effect \citep{LinderWneg1Mirage} for constant-$w$ 
models, we would suggest that constant-$w$ fitting models are {\it not} 
a reliable method for determining the properties of dark energy, 
and should not be used in cosmology at all, if possible. In addition, 
for cosmological analyses assuming any type of $w(z)$ functions, 
we would suggest that model-building uncertainties should always be 
{\it explicitly estimated} by fitting the same data with a variety of 
different DE models, to directly verify that the results do not change 
significantly when the design (or the number) of fitting parameters 
is altered. 

Within the limitations discussed above, it remains interesting to estimate 
just how tight a constraint we can reasonably place upon $j_{0}$ (and thus 
$\Lambda$), with this data. Following \citet{CattVisserDfParamDfFits,
CattVisserYvsZconverge,CattVisserCosmographSNeFits}, the statistical 
justification of a particular fitting function can be put on a 
quantitative footing by considering the ``$F$-test of additional terms'' 
for each case. The improvement in going from a fit (to the 580 SNe data 
points) using $N$ fitting parameters -- with chi-square value 
``$\chi ^{2} _{N}$,'' calculated via Equation~\ref{LstSqrsEqn} -- to one 
with $(N+1)$ parameters (and chi-square $\chi ^{2} _{N+1}$), 
is quantified by computing the statistic:
\begin{equation}
F_{\chi} = 
\frac{\chi ^{2} _{N} - \chi ^{2} _{N+1}}{\chi ^{2} _{N+1}} 
\cdot [580 - (N+1)] ~. 
\label{EqFtestChiSquareStat}
\end{equation}
Since this statistic follows an $F$-distribution with 
$\nu _{1} = 1$ and $\nu _{2} = [580 - (N+1)]$, 
we define the ``$F$-test probability of improvement'' 
(in going from $N$ to $(N+1)$ fitting terms), as: 
\begin{equation}
F_\mathrm{Prob.} = 
[1 - \int^{\infty}_{F_{\chi}} F\{x; 1,[580 - (N+1)]\}~dx] 
\times 100\% ~.
\label{EqFtestProbImprove}
\end{equation}
For each of the fits described above, we calculated 
$\chi ^{2} _{N}$ and used those values to compute $F_{\chi}$ 
(and thus $F_\mathrm{Prob.}$) in going from 
the next-lowest-$N$ case to that one; the resulting 
numbers are given in Table~\ref{TAByRedshiftFtest} for the 
$y$-redshift fits. (Note that these are the exact same 
fits whose cosmological parameters were given above, in 
Table~\ref{TAByRedshiftCosParams}, though here the fits 
are grouped together differently, to better demonstrate 
the effects of increasing $N$.)

\begin{table}
\begin{center}
\caption{\label{TAByRedshiftFtest}
Fit Likelihoods for Expansions in $y$-Redshift, Union2.1 SNe data set.}
{
\begin{tabular}{ccccc}
Fitting Fcn.\ & Fit Terms $N$ & $\chi ^{2}$ 
& $F_{\chi}$ & $F_\mathrm{Prob.}$ \\
\tableline
 & 3 & 563.01 & --- & --- \\
$\ln \{ [ d_\mathrm{L}(y) ] / y \}$ & 4 
     & 562.47 & 0.549 & 54.1$\%$ \\
 & 5 & 561.30 & 1.195 & 72.5$\%$ \\
 & 6 & 561.20 & 0.109 & 25.8$\%$ \\
\tableline
 & 3 & 562.41 & --- & --- \\
$\ln \{ [ d_\mathrm{F}(y) ] / y \}$ & 4 
     & 562.33 & 0.084 & 22.8$\%$ \\
 & 5 & 561.31 & 1.046 & 69.3$\%$ \\
 & 6 & 561.20 & 0.117 & 26.7$\%$ \\
\tableline
 & 3 & 562.23 & --- & --- \\
$\ln \{ [ d_\mathrm{P}(y) ] / y \}$ & 4 
     & 562.20 & 0.026 & 12.8$\%$ \\
 & 5 & 561.32 & 0.906 & 65.8$\%$ \\
 & 6 & 561.20 & 0.125 & 27.6$\%$ \\
\tableline
 & 3 & 562.45 & --- & --- \\
$\ln \{ [ d_\mathrm{Q}(y) ] / y \}$ & 4 
     & 562.09 & 0.374 & 45.9$\%$ \\
 & 5 & 561.33 & 0.776 & 62.1$\%$ \\
 & 6 & 561.20 & 0.134 & 28.5$\%$ \\
\tableline
 & 3 & 563.08 & --- & --- \\
$\ln \{ [ d_\mathrm{A}(y) ] / y \}$ & 4 
     & 561.98 & 1.130 & 71.2$\%$ \\
 & 5 & 561.34 & 0.657 & 58.2$\%$ \\
 & 6 & 561.20 & 0.143 & 29.4$\%$ \\
\tableline
\end{tabular}
}
\end{center}
\end{table}

The lessons from this $F$-testing procedure are somewhat mixed. 
Table~\ref{TAByRedshiftFtest} only indicates a preference 
for $N = 4$ over $N = 3$ about half the time, even though the 
cosmological parameters for $N = 4$ 
(e.g., $w^{4}_\mathrm{X}$ and $j^{4}_\mathrm{X}$ from 
Table~\ref{TAByRedshiftCosParams}) remain physically 
reasonable for all five distance functions. On the other hand, 
we have seen how the best-fit values of the $y$-redshift 
cosmological parameters start undergoing large changes already by 
$N = 5$, yet the $F$-test results in Table~\ref{TAByRedshiftFtest} 
give opposite indications -- that if one has already gone to $N = 4$, 
then going to $N = 5$ would be generally favored.

To sum up the best lessons that can be drawn from this variety of results: 
(i) The $N =5$ and $N = 6$ fits are not useful for producing cosmological 
constraints here, due to very large statistical uncertainties, 
a result echoed (for the $N = 6$ case) by poor performance in the $F$-tests; 
(ii) The estimated cosmological parameters from the $N = 3$ fits are not 
generally reliable, due to large model-building uncertainties which lead to  
best-fit parameter variations (e.g., $\Delta _\mathrm{LA} w ^{3}$, 
$\Delta _\mathrm{LA} j ^{3}$) far in excess of their calculated statistical 
uncertainties; (iii) The $N = 4$ fits are the best compromise for this SNe 
data set, moderating both the statistical and model-building uncertainties, 
while also being mildly preferred by the $F$-tests. (For comparison, note that 
the older SNe data considered in \citet{CattVisserDfParamDfFits,CattVisserYvsZconverge,
CattVisserCosmographSNeFits} -- the Riess Gold06 \citep{RiessGoldSilver} and 
SNLS legacy05 \citep{AstierSNLSleg05} data sets -- were not sufficient 
to go beyond even the $N = 3$ case, which is the bare minimum number of 
parameters needed for estimating $j_{0}$ at all.) 

\begin{figure}
\includegraphics[width=\linewidth]{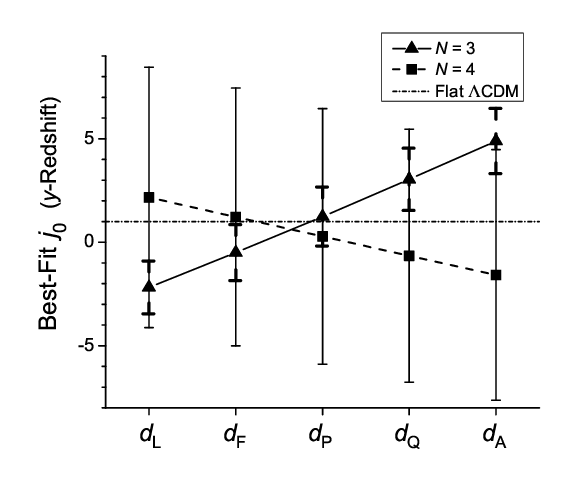}
\caption{Values of Jerk Parameter $j_{0}$ from Union2.1 fits using 
distance functions $\ln \{ [ d_\mathrm{X}(y) ] / y \}$ 
(with $\mathrm{X} \in \{\mathrm{L}, \mathrm{F}, \mathrm{P}, 
\mathrm{Q}, \mathrm{A} \}$), versus the number of series 
terms used, $N = \{ 3, 4 \}$.
\label{Fig2BPD2j0fitsYwithSigs}}
\end{figure}

To illustrate our cosmographic parameter results, the best-fit 
$j_{0}$ values for the $N = \{ 3,4 \}$ $y$-redshift expansions 
are shown in Figure~\ref{Fig2BPD2j0fitsYwithSigs}. The first 
conclusion evident from this plot, is that the $j_{0} = 1$ line 
for flat $\Lambda$CDM is entirely consistent with these results, 
so that there is no statistically meaningful evidence of a 
departure from the cosmological constant. (Interestingly, the 
$N = 3$ and $N = 4$ trends for $j_{0}$ actually cross very close 
to that line, around $j_{0} \sim 0.6$.)

From this figure (and from Table~\ref{TAByRedshiftCosParams}), 
we see that going from $N = 3$ to $N = 4$ cuts the 
model-building uncertainties nearly in half, from 
$\Delta _\mathrm{LA} j ^{3} (y) = 7.07$ to 
$\Delta _\mathrm{LA} j ^{4} (y) = 3.75$. The statistical 
uncertainties $\sigma _{j_{0}}$, however, are increased by 
a factor of about $\sim$$4 - 5$. The $N = 3$ fits would seem 
actually to produce tighter constraints for $j_{0}$ than $N = 4$ -- 
except that it is readily apparent in Figure~\ref{Fig2BPD2j0fitsYwithSigs} 
that the $N = 3$ trend of results has not ``converged'' in terms of 
model-building uncertainty, with the best-fit $j^{3}_\mathrm{X} (y)$ 
numbers changing from one distance-scale fitting function to the next 
by even more than each fit's $\sigma _{j_{0}}$ value. Thus we must clearly 
take the $N = 4$ case as our most reliable source of jerk parameter constraints. 

Therefore, if we (somewhat informally) define the estimated range of 
$j_{0}$ as being bounded by its extreme values -- folding together the 
(1$\sigma$) statistical uncertainties {\it and} model-building uncertainty -- 
we get our ``best possible'' constraint from this cosmographic study: 
$\{ j^{4}_\mathrm{X} (y) \pm \sigma _{j_{0, X}} \} 
\sim [(j^{4}_\mathrm{A} (y) - \sigma _{j_{0, A}}),(j^{4}_\mathrm{L} (y) 
+ \sigma _{j_{0, L}})] 
= [(-1.58 - 6.05),(2.17 + 6.29)] 
= [-7.63,8.46]$; 
or roughly speaking, $j_{0} \sim [-7.6,8.5]$, 
and thus $\Delta j_{0} \sim 16$. One cannot really do better than this 
constraint without the substantial influx of new standard candle data.

\section{\label{SecSimulations}Investigating Model-Building Uncertainties 
with Simulated Supernova Data} 

To estimate the power of cosmographic analysis in the future, when 
many more SN Ia measurements will become available, we construct 
sets of mock supernova magnitude data to study their impact upon  
cosmographic parameter estimation. To avoid making assumptions regarding 
the systematics of the particular surveys that the new SNe data will be 
coming from, we will simply draw our mock SNe data sets from the 
statistical properties of the SCP Union2.1 data set itself. The aim here 
is to ask two basic questions -- what happens when one obtains: 
(a) much more data; (b) higher-redshift data. We therefore construct 
two distinct types of mock data sets, and investigate the best-fit 
parameters obtained from the combination of each of these with the 
original Union2.1 SNe data. 

The first type of mock data set is constructed to double our number of available 
data points. With the Union2.1 SNe data lying within the redshift range, 
$z \in [0.015, 1.414]$ (i.e., $y \in [0.015,0.586]$), 
we construct $58$ $y$-redshift bins from $y = 0.01$ 
to $y = 0.59$, with $\Delta y = 0.01$ per bin. Dividing the $580$ real Union2.1 SNe 
into their appropriate bins, we then generate those same numbers of mock SNe 
in each bin; but rather than make their $y$ values identical to those of Union2.1, 
we randomize the $y$-redshift of each mock SN {\it within} its bin. To produce 
appropriate error bars and scatter, we take the average of the magnitude 
uncertainties for all of the real SNe within a bin, and use that average as the 
value of $\sigma$ for each of the mock SNe in that bin. To generate an actual 
magnitude value for each mock SN, we calculate its ``residual'' via randomization 
from a normal distribution with standard deviation equal to its $\sigma$ value, 
and then add that residual to the appropriate magnitude for a flat $\Lambda$CDM 
model at that $y$-redshift. Performing a simple optimization of flat $\Lambda$CDM 
for the Union2.1 data set yields a minimized value of $\chi^{2} = 562.257$ for 
$H_{0} = 70.0$ km s$^{-1}$ Mpc$^{-1}$ and $\Omega _{\Lambda} = 0.72$, 
which are the fiducial $\Lambda$CDM parameters that we will use for all of our 
mock data generation. 

The construction of a 580-point mock SNe data set can be executed many 
times, with each new realization consisting of completely re-randomized values 
of $y$-redshift and magnitude for each of the 580 simulated data points. 
For each such simulation, the generated mock data set can be added to the real 
Union 2.1 data to produce a 1160-point combined data set, and then subjected to 
the same cosmographic parameter analysis as in Table~\ref{TAByRedshiftCosParams}. 
Running two hundred such simulations, the resulting parameter estimates have been 
averaged over these simulated cases (with the quoted sigmas now being the 
sample standard deviations of the parameters for the 200 trials), and the 
results are given in Table~\ref{TAByRealPlusMock580CosParams}. For brevity, 
we now focus solely upon the equation of state and jerk parameter results 
(henceforth omitting $H_{0}$ and $q_{0}$, as well as the less useful $N = 6$ fits).

\begin{table}
\begin{center}
\caption{\label{TAByRealPlusMock580CosParams}
Cosmographic Parameters from $y$-Redshift fits for the combination of 
Union2.1 SNe plus the statistically similar 580-point mock data set 
($N_{\mathrm{SNe}} = 1160$ total), averaged over 200 mock data realizations.}
{
\begin{tabular}{cccc}
Fitting Function & $N$ & $w_{0}$ & $j_{0}$ \\
\tableline
$\ln \{ [ d_\mathrm{L}(y) ] / y \}$ & & -0.587 & -2.18 ($\pm 0.69$) \\
$\ln \{ [ d_\mathrm{F}(y) ] / y \}$ & & -0.660 & -0.50 ($\pm 0.73$) \\
$\ln \{ [ d_\mathrm{P}(y) ] / y \}$ & 3 & -0.732 & 1.23 ($\pm 0.78$) \\
$\ln \{ [ d_\mathrm{Q}(y) ] / y \}$ & & -0.805 & 3.01 ($\pm 0.82$) \\
$\ln \{ [ d_\mathrm{A}(y) ] / y \}$ & & -0.878 & 4.85 ($\pm 0.86$) \\
                                    & & ($\pm 0.056$) &   ($\dots$) \\
\tableline
$\ln \{ [ d_\mathrm{L}(y) ] / y \}$ & & -0.767 & 2.80 ($\pm 2.91$) \\
$\ln \{ [ d_\mathrm{F}(y) ] / y \}$ & & -0.743 & 1.86 ($\pm 2.89$) \\
$\ln \{ [ d_\mathrm{P}(y) ] / y \}$ & 4 & -0.719 & 0.91 ($\pm 2.87$) \\
$\ln \{ [ d_\mathrm{Q}(y) ] / y \}$ & & -0.695 & -0.02 ($\pm 2.84$) \\
$\ln \{ [ d_\mathrm{A}(y) ] / y \}$ & & -0.671 & -0.95 ($\pm 2.81$) \\
                                    & & ($\pm 0.113$) &   ($\dots$) \\
\tableline
$\ln \{ [ d_\mathrm{L}(y) ] / y \}$ & & -0.470 & -9.52 ($\pm 9.01$) \\
$\ln \{ [ d_\mathrm{F}(y) ] / y \}$ & & -0.477 & -9.12 ($\pm 9.03$) \\
$\ln \{ [ d_\mathrm{P}(y) ] / y \}$ & 5 & -0.484 & -8.72 ($\pm 9.05$) \\
$\ln \{ [ d_\mathrm{Q}(y) ] / y \}$ & & -0.492 & -8.33 ($\pm 9.07$) \\
$\ln \{ [ d_\mathrm{A}(y) ] / y \}$ & & -0.499 & -7.93 ($\pm 9.08$) \\
                                    & & ($\pm 0.241$) &   ($\dots$) \\
\tableline
\end{tabular}
}
\end{center}
\end{table}

The parameter values in Table~\ref{TAByRealPlusMock580CosParams} should be 
compared to those from Table~\ref{TAByRedshiftCosParams} for the real 
Union2.1 SNe data set alone. First, we see that adding a simulated data set 
(derived directly from flat $\Lambda$CDM with explicitly gaussian variations) 
cuts the statistical uncertainties roughly in half. The fits with low $N$, 
which have few degrees of freedom and small statistical uncertainties anyway, 
experience only minor changes to their best-fit $w^{N}_\mathrm{X}$ and 
$j^{N}_\mathrm{X}$ values; but for higher $N$ cases (particularly $N = 5$ here), 
there is a stronger ``corrective'' effect, dragging the fits much closer 
(and with smaller sigmas) to the imposed mock values of $w_{0} = -0.72$ 
and $j_{0} = 1$.

Interestingly, adding of all this extra ``data'' (with the {\it same} redshift 
distribution as Union2.1) has almost no effect on the size of the 
model-building uncertainties, $\Delta _\mathrm{LA} w ^{N}_\mathrm{X} (y)$ and 
$\Delta _\mathrm{LA} j ^{N}_\mathrm{X} (y)$, when one compares the results in 
Tables~\ref{TAByRedshiftCosParams},\ref{TAByRealPlusMock580CosParams} for the 
{\it same} $N$. But what adding extra data does in fact accomplish, is that by 
reducing the statistical uncertainties on all of the estimated parameters, 
it becomes more feasible to now use fits with higher $N$ (more fitting parameters), 
which leads naturally to smaller model-building uncertainties. For example, 
the sigmas on $w^{N}_\mathrm{X}$ for the $N = \{4, 5 \}$ cases with the 
mock data are now roughly as small as they had been for the 
$N = \{3, 4 \}$ cases with the Union2.1 data alone. Folding together 
both the statistical and model-building uncertainties as before, 
we see (for example) that the $N = 3$ case for Union2.1 alone only allowed us to 
constrain the observed EoS within $w^{\mathrm{Obs}}_{0} \sim [-0.97,-0.48]$; 
but the $N = 4$ case (for Union2.1 plus mock data) now allows us to constrain 
$w^{\mathrm{Obs}}_{0} \sim [-0.88,-0.56]$, a $\sim$35$\%$ narrowing of the allowed range. 

For the crucial (and harder to measure) jerk parameter, the statistical uncertainty 
reductions are not yet enough to allow us to usefully move up to a higher $N$; 
but even staying within $N = 4$, adding the mock data allows us to narrow the original 
range by $\sim$40$\%$, from $j_{0} \sim [-7.6,8.5]$ to $j_{0} \sim [-3.8,5.7]$. 
And adding just a few hundred more SNe Ia of similar quality would eventually 
reduce the importance of the statistical sigmas by enough (compared to the 
model-building uncertainties) to allow us to move from $N = 4$ up to $N = 5$, 
with the resulting improvement from 
$\Delta _\mathrm{LA} j ^{4}_\mathrm{X} (y) \sim 3.8$ to 
$\Delta _\mathrm{LA} j ^{5}_\mathrm{X} (y) \sim 1.6$.

Now consider the generation of a higher-redshift mock data set, 
hypothetically capable of constraining higher-order terms in the 
distance scale function Taylor expansions. With the goal of exploring how 
cosmographic methods generically respond to high-$z$ SNe data, we extrapolate 
these new mock SNe from the high-redshift end of the Union2.1 data set. 

The number of SNe per $y$-redshift bin in Union2.1 begins to fall off 
rapidly past $y \sim 0.51$ ($z \gtrsim 1.04$). To make up for this abrupt 
fall-off -- by filling out the distribution toward its high-redshift end 
($z \simeq 1.4$), and continuing beyond (out to $z \sim 2$ or so) -- 
would require $\sim$$2-3$ dozen mock SNe. This is comparable to 
the $28$ $z > 1$ SNe in the ``HST+WFC3 6yr'' mock sample of 
\citet{SalzanoRefRecFits}; but unfortunately, only $12$ of their 
mock SNe lay significantly beyond the edge of Union2.1 ($z \gtrsim 1.5$), 
and a mere dozen new highest-redshift SNe are not enough to 
change the results determined by the 580 SNe of Union2.1 very much. 

Requiring a larger sample, we generate three dozen mock 
SNe past the Union2.1 limit of $z = 1.414$ ($y \sim 0.59$). Scaling up the 
distribution of our 36 mock highest-redshift SNe from the ratios in 
\citet{SalzanoRefRecFits}, we place $30$ SNe within $z \sim [1.5,2.0]$, 
and $6$ SNe within $z \sim [2.0,2.5]$. More specifically, beyond the edge 
of Union2.1 we create $8$ new $y$-redshift bins covering $0.59 \le y \le 0.67$ 
to contain the first $30$ of these new SNe, and another $4$ bins covering 
$0.67 \le y \le 0.71$ to contain the final $6$ new SNe. While the $y$ 
value of each mock SN is randomized {\it within} its $\Delta y = 0.01$ bin, 
the SNe are divided as evenly as possible among the different bins, to 
simulate the addition of $36$ well-distributed SNe capable of effectively 
tracing out the entire redshift range out to $y = 0.71$ ($z \sim 2.45$). 

To assign error bars for the mock data, we used information from Union2.1, 
with the sigmas designed to get progressively worse for higher redshift. 
For each $y$-bin of the original Union2.1 data, to smooth out the inherent 
variability of the individual SN error bars, we take the average of the 
$\sigma _{\ln [ d_\mathrm{L} ]}$ values for all of the SNe within that bin 
(call this bin-averaged error bar ``$\sigma ^{\mathrm{bin}} _{i}$,'' for the 
$i^\mathrm{th}$ bin). Examining these $\sigma ^{\mathrm{bin}} _{i}$ values 
shows that they tend to increase sharply up to $y \sim 0.27$ ($z \sim 0.37$), 
after which they appear to vary roughly normally around an average of 
$\langle \sigma ^{\mathrm{bin}} _{i} \rangle = 0.135$ for the $32$ remaining 
bins from $0.27 < y < 0.59$, with a sample standard deviation of 
$\sigma \{ \sigma ^{\mathrm{bin}} _{i} \} = 0.023$. For the 
$30$ high-redshift mock SNe within $y \in [0.59,0.67]$, to produce sigmas 
larger than the Union2.1 average (but by a statistically realistic amount), 
we set their error bars equal to $\sigma _{\ln [ d_\mathrm{L} ]} \equiv 
\langle \sigma ^{\mathrm{bin}} _{i} \rangle 
+ \sigma \{ \sigma ^{\mathrm{bin}} _{i} \}  = 0.158$. 
Then, for the $6$ highest-redshift mock SNe within $y \in [0.67,0.71]$, 
to get the largest reasonable uncertainties (based upon Union2.1), we used 
the highest bin-averaged uncertainty from Union2.1, so that 
$\sigma _{\ln [ d_\mathrm{L} ]} 
\equiv \mathrm{Max} \{ \sigma ^{\mathrm{bin}} _{i} \} = 0.177$. 

Finally, as was done with the 580-point mock data set, the magnitudes 
of these simulated SNe are based upon a flat $\Lambda$CDM model 
with $H_{0} = 70.0$ km s$^{-1}$ Mpc$^{-1}$, $\Omega _{\Lambda} = 0.72$, 
with residuals generated randomly from normal distributions with 
standard deviations equal to the sigma values discussed above.
Once again, 200 simulations were run, generating mock data; and as 
the highest-redshift data set contains only a small number of SNe 
on its own, each mock data set was added to the real Union2.1 
data to form a combined set of $616$ SNe, which was then 
subjected to cosmographic analysis. The averaged output parameters 
(with sample standard deviations) of these simulations 
are given in Table~\ref{TAByCosParHiRedMockPlusUn21}.

\begin{table}
\begin{center}
\caption{\label{TAByCosParHiRedMockPlusUn21}
Cosmographic Parameters from $y$-Redshift fits for the 
combination of Union2.1 SNe plus the highest-redshift ($y > 0.59$) 
mock data set ($N_{\mathrm{SNe}} = 616$ total), 
averaged over 200 mock data realizations.}
{
\begin{tabular}{cccc}
Fitting Function & $N$ & $w_{0}$ & $j_{0}$ \\
\tableline
$\ln \{ [ d_\mathrm{L}(y) ] / y \}$ & & -0.516 & -3.20 ($\pm 0.54$) \\
$\ln \{ [ d_\mathrm{F}(y) ] / y \}$ & & -0.629 & -0.99 ($\pm 0.59$) \\
$\ln \{ [ d_\mathrm{P}(y) ] / y \}$ & 3 & -0.741 & 1.35 ($\pm 0.63$) \\
$\ln \{ [ d_\mathrm{Q}(y) ] / y \}$ & & -0.854 & 3.81 ($\pm 0.67$) \\
$\ln \{ [ d_\mathrm{A}(y) ] / y \}$ & & -0.967 & 6.40 ($\pm 0.72$) \\
                                    & & ($\pm 0.039$) &   ($\dots$) \\
\tableline
$\ln \{ [ d_\mathrm{L}(y) ] / y \}$ & & -0.800 & 3.75 ($\pm 2.10$) \\
$\ln \{ [ d_\mathrm{F}(y) ] / y \}$ & & -0.754 & 2.13 ($\pm 2.06$) \\
$\ln \{ [ d_\mathrm{P}(y) ] / y \}$ & 4 & -0.708 & 0.53 ($\pm 2.03$) \\
$\ln \{ [ d_\mathrm{Q}(y) ] / y \}$ & & -0.662 & -1.06 ($\pm 2.00$) \\
$\ln \{ [ d_\mathrm{A}(y) ] / y \}$ & & -0.615 & -2.62 ($\pm 1.97$) \\
                                    & & ($\pm 0.068$) &   ($\dots$) \\
\tableline
$\ln \{ [ d_\mathrm{L}(y) ] / y \}$ & & -0.466 & -8.94 ($\pm 5.04$) \\
$\ln \{ [ d_\mathrm{F}(y) ] / y \}$ & & -0.485 & -8.03 ($\pm 5.07$) \\
$\ln \{ [ d_\mathrm{P}(y) ] / y \}$ & 5 & -0.505 & -7.11 ($\pm 5.09$) \\
$\ln \{ [ d_\mathrm{Q}(y) ] / y \}$ & & -0.524 & -6.18 ($\pm 5.12$) \\
$\ln \{ [ d_\mathrm{A}(y) ] / y \}$ & & -0.544 & -5.26 ($\pm 5.14$) \\
                                    & & ($\pm 0.123$) &   ($\dots$) \\
\tableline
\end{tabular}
}
\end{center}
\end{table}

It is interesting to compare the numbers in 
Table~\ref{TAByCosParHiRedMockPlusUn21} for this highest-redshift 
mock data set added to Union2.1, versus Table~\ref{TAByRedshiftCosParams} 
(for the Union2.1 data alone), and Table~\ref{TAByRealPlusMock580CosParams} 
(with the 580-point medium-redshift mock data set added to Union2.1). 
One first notices that the addition of a mere three dozen SNe at very high 
redshift does a similarly good job of ``correcting'' the parameters toward 
the (imposed mock) values of $w_{0} = -0.72$ and $j_{0} = 1$, while 
simultaneously reducing the parameter statistical uncertainties even more 
than was done by adding in the much larger set of medium-redshift mock data. 

On the other hand, the statistical uncertainties have been reduced 
{\it so much}, that a number of the best-fit parameters (particularly 
for the $N = 5$ case) are no longer within 1$\sigma$ of $w_{0} = -0.72$, 
$j_{0} = 1$. A related problem with these best-fit cosmographic parameters, 
is that the addition of the higher-redshift data has made the model-building 
uncertainties {\it worse}, increasing $\{ \Delta _\mathrm{LA} j ^{3} (y), 
\Delta _\mathrm{LA} j ^{3} (y), \Delta _\mathrm{LA} j ^{3} (y) \}$ from 
$\{ 7.07, 3.75, 1.57 \}$ (for Union2.1 data alone) to $\{ 9.60, 6.37, 3.68 \}$, 
with similarly serious increases to the values of 
$\Delta _\mathrm{LA} w ^{N} (y)$. Thus even though the constraints on 
$\{ w_{0}, j_{0} \}$ might appear to be tighter due to the reduction of the 
statistical uncertainties, they are actually significantly less robust 
due to the increased model-building uncertainties. And since no 
comparable detrimental effect was seen when adding in the 580-point 
mock data set -- which had the same redshift distribution as 
Union2.1 -- it is clear that the problem is specifically due to the 
high-redshift nature of these new mock SNe.

The obvious source of this difficulty is the same ``fitting a line 
to a parabola'' problem explained in Section~\ref{SecIntro}. 
As new data constrains the fits at higher and higher redshifts, 
high-order terms in the Taylor series expansions become larger 
and increasingly non-negligible. Thus any truncated 
(i.e., finite-term) series fit using the {\it same} number of 
terms $N$ will become more model-dependent as the data fills out 
a larger redshift range, without the fitting model growing in 
complexity to match it. Therefore, adding high-redshift data 
(of any kind) 
actually makes the estimation of the three lowest-order parameters 
($H_{0}$, $w_{0}$, and $j_{0}$) {\it worse} for cosmography (and for 
any other fitting method using too few optimizable parameters to 
accurately trace out the curve followed by the data). 

Since it is imperative to obtain parameter estimates that are reliably 
model-independent, these model-building uncertainties must be reduced. 
The two obvious ways to do this, are to either the increase the number 
$N$ of fitted parameters -- which in all cases increases the statistical 
uncertainties by a great amount -- or to limit the redshift range of the 
data being retained for the fits. In BPDv2, we show that an effective way 
to reduce the cosmographic model-building uncertainties is by actually 
truncating the data set above a certain redshift cutoff; we used $z \sim 1$, 
$y \sim 0.5$, yielding a $\sim$$20 - 60 \%$ reduction for various cases. 
Though it is always unfortunate to discard real (and hard to obtain) 
high-redshift data -- and though doing so {\it also} leads to some increase 
in the parameter statistical uncertainties -- such statistical sigmas can 
always be reduced via the acquisition of more (medium-redshift) data; 
but there is no amount of data that can reduce the model-building 
uncertainties for models with an inadequate number $N$ of fitting terms. 

Recalling that the flat $\Lambda$CDM paradigm can be falsifiably 
tested by obtaining the value of $j_{0}$ from just (two of) the three 
lowest-order terms in the Taylor series expansion of any of the 
distance scale functions (e.g., from $p_{1}$ and $p_{2}$ via 
Equations~\ref{CosmoParamq0Eqn},\ref{CosmoParamj0Eqn}), the retention 
of high-redshift data (and higher-order Taylor series terms) is 
actually counter-productive toward that highly specific goal. In essence, 
a redshift-limited cosmographic analysis like the one we suggest just 
represents the logical extension of historical attempts to obtain a 
precise measurement of the Hubble Constant by restricting the analysis 
of SNe Ia to use only those data lying below the redshift where the 
linear ``Hubble Law'' begins to break down.

\section{\label{SecDiscConclusion}Conclusions}

In this paper, we have presented the results of a cosmographic 
study designed to test for deviations from the theoretically simplest 
accelerating universe model -- flat $\Lambda$CDM, containing only matter 
and a cosmological constant -- by searching for tell-tale deviations of 
the jerk parameter from $j_{0} = 1$. For this purpose, we have utilized 
the data set most useful for continuously tracing out the evolution of the 
universe over the acceleration epoch: Type Ia supernova standard candles, 
using magnitude data available from the SCP Union2.1 compilation. 
In the process, we have also studied the cosmographic analysis method 
itself, using mock data extrapolated from Union2.1 to evaluate its 
capability for cosmological parameter estimation in future studies. 

Estimates of the Hubble Constant $H_{0}$, the cosmic total 
equation of state $w_{0}$ (from deceleration parameter $q_{0}$), 
and $j_{0}$ were derived from the data by performing polynomial 
fits, based upon Hubble series expansions in powers of the cosmological 
redshift. The results quoted here were done using expansions in the 
newer $y$-redshift variable, $y \equiv z/(1+z)$, due to its reliable 
series convergence properties.

Any fitting process utilizing a finite number of optimizable parameters 
will have results that depend to some degree upon the particular model 
(i.e., the type and number of such parameters) being used; 
in the context of cosmography, this indeterminacy has been 
referred to as ``model-building uncertainty''. To test whether 
the cosmological parameter estimates from data sets such 
as this one are still jeopardized by model-building uncertainties 
(relative to the limitations due to statistical uncertainties), we 
performed fits using convenient logarithmic forms of five different 
distance indicator functions: the luminosity distance ($d_\mathrm{L}$), 
the photon flux distance ($d_\mathrm{F}$), the photon count distance 
($d_\mathrm{P}$), the deceleration distance ($d_\mathrm{Q}$), and the 
angular diameter distance ($d_\mathrm{A}$). Furthermore, we performed 
fits using four different degrees of polynomial series expansion in each 
case, extracting a different number (``$N$'') of optimized coefficients 
each time -- going up to $N = \{ 3, 4, 5, 6 \}$ 
(i.e., $\{ O[y^{2}], \dots , O[y^{5}] \}$) -- 
in order to test the balance between smaller model-building uncertainty 
for higher $N$ (due to reduced series truncation errors), versus higher 
statistical uncertainty in the fit coefficients (and thus implausible 
cosmological parameter estimates) for values of $N$ larger than the 
number of measurable quantities that can be reliably extracted from 
the current SNe data. 

Our first main conclusion, is that one cannot reliably extract more 
than four meaningful cosmographic parameters from the Union2.1 supernova 
compilation. Fits going beyond $O[y^{3}]$ become visibly unconstrained 
as soon as the SNe data runs out at high redshift, around $y \gtrsim 0.58$ 
($z \gtrsim 1.4$). Consequently, their best-fit cosmological 
parameters become quite implausible compared to typical estimates from 
the literature (though they still typically lie within 1$\sigma$ of them, 
given the very large statistical uncertainties for parameters from 
the $N = \{5, 6 \}$ fits). 

In theory, the ability to extract four Hubble series parameters 
from the data would appear to be sufficient to not only accurately 
estimate the crucial jerk parameter, $j_{0}$ (along with $H_{0}$ 
and $q_{0}$), but also to estimate the snap parameter, $s_{0}$. 
Unfortunately, even going beyond $N = 3$ to $N = 4$ terms 
is not enough to reliably disentangle estimates of $j_{0}$ from 
the model-to-model variation due to series truncation error. 
Thus for the $N = 4$ case, the $y$-redshift fits only constrain 
the jerk parameter within the broad range of $j_{0} \sim [-7.6,8.5]$, 
when statistical uncertainties $\sigma _{j_{0}}$ are folded in along 
with those model-building uncertainties. Flat $\Lambda$CDM is entirely 
consistent with this result, but it is not a very strong constraint, 
with $\Delta j_{0} \sim 16$. 

To estimate the benefits obtainable with more data from future surveys, 
we used the statistical properties of the Union2.1 compilation to 
construct two types of mock data sets: a 580-point set of simulated SNe 
with a similar redshift distribution as Union2.1; and a high-redshift 
data set with 36 mock SNe spread over the range $y \in [0.59,0.71]$ 
(i.e., $z \sim 1.5 - 2.5$). Each type of mock data set was 
simulated 200 times, with the cosmographic parameter results from each 
of the separate, randomized realizations averaged together afterwards, 
for each of the two classes of mock data set.

Combining the 580-point mock data set with Union2.1, a cosmographic 
analysis indeed shows the kind of reduction expected in the 
statistical uncertainties, so that this experiment of ``doubling'' 
the available supernova data successfully tightens the constraints 
to $j_{0} \sim [-3.8,5.7]$, or $\Delta j_{0} \sim 9.5$. Crucially, 
however, adding this extra data -- which possesses the same redshift 
distribution as the original Union2.1 data set -- makes virtually 
no change in the model-building uncertainties, when comparing fits 
done using the {\it same number} of fitting parameters, $N$.

Incorporating the high-redshift mock data set (in combination 
with Union2.1), however, makes the model-building uncertainty situation 
significantly {\it worse}, causing a substantial increase in how much 
the cosmological parameters change (for a given polynomial order $N$) 
when cycling through the distance scale functions 
$\{ d_\mathrm{L}, d_\mathrm{F}, d_\mathrm{P}, d_\mathrm{Q}, d_\mathrm{A} \}$. 
This happens because the truncated higher-order polynomial terms 
become more important at higher redshift, forcing the lower-order 
terms to adjust (inappropriately) to compensate for their absence. 
This leads to an increased variation in the estimated cosmological 
parameters, $\{ H_{0}, w_{0}, j_{0} \}$, which are drawn from just 
the first three series terms in each fit. 

Since simply adding more data does not help reduce these 
model-building uncertainties, one could perhaps go to higher-order 
polynomial fits (i.e., larger $N$); though we show that this leads 
to a huge increase in the statistical uncertainties on {\it all} of 
the best-fit parameters, more than negating the benefits of the 
reductions obtained for the model-building uncertainties. 

Alternatively, one can make the high-order Taylor series terms 
unimportant by reducing the redshift range of the data, 
essentially truncating the standard candle data sets used for the 
fits above some chosen redshift cutoff (e.g., $z \gtrsim 1$). While 
discarding real data is a sacrifice, and one which also increases 
the statistical uncertainties to some degree (though not by as much 
as when forced to go higher $N$), if the main goal is to apply a 
falsifying test for the cosmological constant, then the only quantity 
of interest (assuming a flat universe) is the jerk parameter, 
which can be effectively specified by just the first 3 terms in the 
Taylor series expansions. Thus the very difficult task of acquiring 
high-redshift, high-precision standard candle data becomes superfluous, 
and the most straightforward strategy for testing $\Lambda$ 
is just the acquisition of enough moderate-redshift 
(say, $z \sim 0.1 - 1$) standard candle data to ultimately beat down 
the statistical uncertainties to very small levels for the 
$N = 3$ or $N = 4$ (or ultimately even $N = 5$) cosmographic fits.

Lastly, perhaps the most important lesson from this paper is that 
an essential aspect of cosmological parameter estimation is the 
performance of explicit tests to verify the {\it model independence} 
of the results. This is done by using different best-fit functions 
which vary the form and number of the optimized parameters, in order 
to confirm that the model-to-model variations in the results are small 
compared to the other (statistical, systematic) sources of uncertainty. 
For dynamical dark energy fits, this obviously implies the use of different 
types of equation of state functions, $w(z)$, verifying similar results 
before any parameter estimates are quoted. But for cosmographic studies, 
specifically, we conclude that the parameter estimation results 
{\it must} be compared using different distance scale functions -- 
say, luminosity distance $d_\mathrm{L}$ versus angular diameter distance 
$d_\mathrm{A}$ -- for whatever polynomial order(s) $N$ are being used for 
the quoted fits results, to demonstrate the smallness of the 
model-building uncertainties. If this is not done, then the best-fit 
cosmographic parameters which result cannot be considered to be 
robust estimates, no matter how much care has been taken with all 
other aspects of the overall error budget.

\acknowledgments

We are grateful to Marie van Nieuwenhuizen for her academic support 
of one of the authors (M. D.) during the course of this research.

\end{document}